\documentclass[aps,twocolumn,preprintnumbers]{revtex4}
\usepackage{mathrsfs}
\usepackage{amsmath, amssymb, amsfonts, bbm}
\usepackage{graphics,epsfig,color,subfigure,graphicx,epsf}

\newcommand{\be}{\begin{equation}}
\newcommand{\ee}{\end{equation}}
\def\bea{\begin{align}}
\def\ena{\end{align}}

\def\nnb{\nonumber}

\def\bea{\begin{eqnarray}}
\def\eea{\end{eqnarray}}
\def\nnb{\nonumber}
\def \Nc{N_c}

\begin{document}
\preprint{UMD-40762-452}
\preprint{PUPT-2300}

\title{Universal relations of transport coefficients from holography}

\author{Aleksey Cherman}
\email{alekseyc@physics.umd.edu}
\affiliation{Center for Fundamental Physics, Department of Physics, University of Maryland,
College Park, MD 20742-4111}

\author{Abhinav Nellore}
\email{anellore@princeton.edu}
\affiliation{Joseph Henry Laboratories, Princeton University, Princeton, NJ 08544}

\begin{abstract}
We show that there are universal high-temperature relations for transport coefficients of plasmas described by a wide class of field theories with gravity duals.  These theories can be viewed as strongly coupled large-$N_c$ conformal field theories deformed by one or more relevant operators.  The transport coefficients we study are the speed of sound and bulk viscosity, as well as the conductivity, diffusion coefficient, and charge susceptibility of probe $U(1)$ charges.  We show that the sound bound $v_s^2 \le 1/3$ is satisfied at high temperatures in these theories and also discuss bounds on the diffusion coefficient, the conductivity and the bulk viscosity.
\end{abstract}

\maketitle

The gauge/gravity duality relates certain strongly coupled large-$\Nc$ quantum field theories to higher-dimensional classical gravity theories~\cite{GaugeGravity}.  In particular, it allows us to study the hydrodynamics of systems described by gauge theories at strong coupling.  Hydrodynamics can be thought of as an effective field theory describing the physics of fluids on distances and time scales that are large compared to the underlying microscopic scales.   Transport coefficients are essentially the low-energy constants of hydrodynamics, and they generally appear as poles of real-time Green's functions of conserved currents.   The transport coefficients of strongly coupled fluids are not easily calculable by theoretical methods other than the gauge/gravity duality.  Where it applies, the duality provides a dictionary for relating the correlation functions of operators in a gauge theory to certain computations in classical gravity.  The detailed prescription for calculating real-time correlation functions was worked out in Refs.~\cite{AdSCFTHydro}.

Unfortunately, since there are no known gravity duals for the gauge theories that are currently used to describe nature, the duality cannot yet be used to make quantitative predictions.  However, we can search for `universal' properties of transport coefficients in strongly coupled large-$\Nc$ theories with gravity duals.  In the process, we may learn some general lessons about hydrodynamics, perhaps ones that apply even to theories without gravity duals.  For instance, the ratio of shear viscosity to entropy density $\eta/s$ takes the value $1/4\pi$ in all theories with Einstein gravity duals~\footnote{The phenomenological implications of the universality of $\eta/s=1/4\pi$ in strongly coupled large $\Nc$ theories are unclear~\cite{ManySpecies}; the corrections to $\eta/s$ away from the infinite-coupling and infinite-$\Nc$ limits do not appear to be universal~\cite{EtaBoundViolations}.}.  So $\eta/s$ is completely independent of the details of these theories~\cite{StretchedHorizons,EtaBound}.  In particular, $\eta/s$ does not depend on the temperature $T$.  A natural next question is: Are there any universal properties of other transport coefficients, which generally \emph{do} depend on temperature?

In the present paper, we answer this question in the affirmative.  We show that a number of transport coefficients have the same universal high-temperature behavior in a broad class of field theories with gravity duals.  We study four-dimensional ($4D$) field theories whose gravity duals contain $n$ scalar fields $\phi_i$, $i=1,\ldots, n$.  The scalar fields have an interaction potential that is assumed to meet some technical conditions, but is otherwise arbitrary.  The field theories can be thought of as strongly coupled large-$\Nc$ $\mathcal{N}=4$ super Yang-Mills theory (SYM) deformed by $n$ relevant operators $\mathcal{O}_{\phi_i}$ with scaling dimensions $\Delta_i$ dual to the bulk fields $\phi_i$.  We review and extend the computation of the speed of sound $v_s$ from Ref.~\cite{vs2CCN}, where some of the techniques we use were developed.  Next, we calculate the diffusion coefficient $D$, the charge susceptibility $\Xi$, and the DC conductivity $\sigma$ of $U(1)$ charges in the probe limit at high temperatures.   We also calculate the bulk viscosity $\zeta$ at high temperatures in single-scalar backgrounds.  A high-temperature calculation of $\zeta$ in theories with gravity duals sourced by multiple scalar fields is left for future work.

Specifically, consider the set of normalized transport coefficients $\xi_i \in \{\zeta/s, v_s^2, 2 \pi T D, \sigma/\pi T, \Xi/(2 \pi^2 T^2) \}$.  By `normalized,' we mean that if we added back appropriate factors of $\hbar$, the speed of light $c$, and the Boltzmann constant $k_B$, these quantities would be dimensionless.  We demonstrate that
\be
\label{universal_relation}
\lim_{T\rightarrow \infty} \frac{d \xi_i}{d \xi_j } = \lim_{T\rightarrow \infty} \frac{\partial \xi_i /\partial \log T}{\partial \xi_j /\partial \log T} = R_{\xi_{i}, \xi_j}(\Delta)  \;,
\ee
where $R_{\xi_{i}, \xi_j}(\Delta)$ is a matrix of \emph{nonzero} constants that depends on $\Delta \equiv \max(\Delta_i)$.   $R_{\xi_{i}, \xi_j}(\Delta)$ does \emph{not} depend on any details of the scalar potential except through $\Delta$.  

The relation above follows from the fact that as $T\rightarrow \infty$, the $\xi_i$ approach their values in $\mathcal{N}=4$ SYM, a conformal field theory (CFT), in the same way.  At high temperatures, we show that
\be
\label{TransportHighT}
\xi_{i}(T) = \xi_{i}^{\mathrm{CFT}} + \mathcal{C}_{\xi_i} (\Delta) T^{-2 (4-\Delta)}
\ee
plus corrections that are suppressed as $T\rightarrow\infty$.  Here, we have defined $\xi_{i}^{\mathrm{CFT}} = \xi_{i}(T\rightarrow \infty)$. 

We always work in units where $\hbar = c = k_B = 1$, and we focus on $4D$ field theories for concreteness.  The organization of the rest of this paper is as follows.  We introduce single-scalar models in Sec.~\ref{sec:SingleScalar} and describe the high-temperature expansion that applies to these models in Sec.~\ref{sec:HighT}.  The calculation of transport coefficients at high $T$ is the focus of Sec.~\ref{sec:TransportCoeffs}.  We discuss the extension of our results to systems with gravity duals sourced by multiple scalar fields in Sec.~\ref{sec:MultiScalar}, and we conclude in Sec.~\ref{sec:Conclusions}.


\section{A single-scalar model}  
\label{sec:SingleScalar}

The simplest class of nonconformal gravity duals is characterized by a `single-scalar' model with the action~\cite{GubserNellore,GursoyEtAl}
\be
\label{S5D}
S = \frac{1}{2 \kappa_{5}^2} \int{d^5\,x \sqrt{-g} \left[ R - \frac{1}{2}(\partial \phi)^2 - V(\phi) \right]} \;,
\ee
where $\kappa_5^2/8 \pi$ is the Newton constant, $\phi$ is a real scalar field, and $V(\phi)$ is an analytic potential.  Moreover, $V(\phi)$ is symmetric about an extremum at $\phi=0$, and $V(0)<0$.  We study finite-temperature field theories that are rotationally invariant in space and translationally invariant in spacetime.  This translates to the metric ansatz
\be
\label{metricansatz}
ds^2 = a^2 (-h dt^2 + d\vec{x}^2)+ \frac{dr^2}{b^2 h}  \;,
\ee
where $a$, $b$, and $h$ are smooth functions of the holographic coordinate $r$ only, and $\phi=\phi(r)$.  A black hole horizon occurs at $r=R_h$, where $h$ has a simple zero.  The entropy density $s$ and temperature $T$ of the field theory are simply the entropy density and temperature of the black hole: 
\be
\label{SandT}
s = \frac{2\pi}{\kappa_{5}^2} |a(R_h)|^3  \, ,\; \; \; \; \;
T = \frac{|a(R_h)b(R_h)h'(R_h)|}{4\pi} \; .
\ee

Different choices of the potential $V$ correspond to different dual gauge theories.  Unlike QCD, these theories are both strongly coupled and conformal in the high-temperature limit.  Their dual geometries approach five-dimensional anti-de Sitter space ($AdS_5$) in the extreme UV, where $\phi\rightarrow 0$.  Our analysis thus relies on the $AdS$/CFT dictionary~\cite{GaugeGravity}.  Formally, the scalar deforms $\mathcal{N}=4$ SYM, a CFT, by adding a term $\Lambda^{4-\Delta}\mathcal{O}_{\phi}$ to its lagrangian.  $\mathcal{O}_{\phi}$ is the field theory operator dual to $\phi$, $\Delta$ is the UV scaling dimension of $\mathcal{O}_{\phi}$, and $\Lambda$ is a new energy scale introduced by the scalar.

At small $\phi$, the potential takes the form
\be
\label{asympV}
V(\phi) = -\frac{12}{L^2} + \frac{1}{2L^2} \Delta(\Delta-4) \phi^2 + \mathcal{O}(\phi^{4}) \;,
\ee
where $L$ is the curvature radius of asymptotic $AdS_5$.   So the squared mass $m^2$ of the scalar is $\Delta(\Delta-4)/L^2$.  $m^2 L^2 \geq -4$ accommodates the Breitenlohner-Freedman stability bound for a scalar in $AdS_5$~\cite{BFbound}.  The allowed range of the scaling dimension $\Delta$ is also constrained.  $\Delta<4$ makes $\mathcal{O}_{\phi}$ a relevant operator, but $\Delta<1$ violates the unitarity bound.   For convenience, we focus on $2< \Delta<4$.  The extension to the case $\Delta \le 2$ is straightforward~\cite{KlebanovWitten}.

\section{High-temperature expansion}
\label{sec:HighT}

In this section, we pursue a high-temperature expansion of the background and scalar profile in single-scalar models.  A similar expansion was developed in the study of $\mathcal{N}=2^*$ theory~\cite{N2SYM}.

Let us work in the gauge $a=r$ with the $AdS$ boundary at $r=\infty$.  Three independent equations of motion follow from the action Eq.~(\ref{S5D}) in the background Eq.~(\ref{metricansatz}):
\be \label{theeoms}\begin{split}
&0 = \frac{2 V(\phi)}{b^2}+\frac{6(4 h  + r h' )}{r^2}-h (\phi')^2\\
&0 =6 r b'+b (-6+r^2 (\phi')^2)\\
&0 = - \frac{V'(\phi)}{b^2}+h' \phi' + h\left((\frac{4}{r}+ \frac{b'}{b})\phi'+\phi'' \right).
\end{split}\ee
The first two come from recombining the $tt$ and $rr$ components of Einstein's equation, and the last is the scalar equation of motion.  When $T\gg\Lambda$, the influence of $\mathcal{O}_{\phi}$ on field theory phenomena becomes negligible.  On the gravity side, this means that the scalar vanishes in the high-temperature limit \footnote{For some $V$, there is a secondary high-temperature limit that corresponds to an extremal black hole.  However, this limit always occurs along a branch of the equation of state that violates the first law of thermodynamics, so we exclude it from consideration.}.  When $\phi$ is everywhere zero, the general solution to Eqs.~(\ref{theeoms}) is $AdS_{5}$-Schwarzschild:
\be \label{adsschwarz}
b = \frac{r}{L} \, ,\; \; \; \; \;
h = 1 - \left(\frac{r_h}{r}\right)^4.
\ee
Above, $r_h$ is the location of the $AdS_{5}$-Schwarzschild black hole horizon.  The entropy density and temperature are computed from Eq.~(\ref{SandT}) as
\be 
\label{adsschwarzsandt}
s = \frac{2\pi}{\kappa_{5}^2} r_{h}^3  \, ,\; \; \; \; \;
T = \frac{r_h}{\pi L} \; .
\ee
Large $r_h$ thus corresponds to large $s$ and $T$, and $s\sim T^3$, as one expects for a $3+1$-dimensional CFT.

When $\phi$ is everywhere small, it takes the form
\be
\label{firstscalar}
\phi(r) = \phi_0 \,{}_{2}F_{1}(1-\Delta/4, \Delta/4;1;1-r^4/r_h^4) \;,
\ee
where $\phi_0$ is an integration constant that measures the smallness of the scalar. Equation~(\ref{firstscalar}) is one solution to the linearized scalar equation of motion with $b$ and $h$ as in Eq.~(\ref{adsschwarz}).  We have discarded the other solution, which diverges logarithmically as $r\rightarrow r_h$.  Note that only the universal terms of the potential shown in Eq.~(\ref{asympV}) determine Eq.~(\ref{firstscalar}).

We would like to compute thermodynamic quantities and transport coefficients at high temperatures, when the scalar is small.  To this end, we develop perturbation expansions of the metric and scalar in powers of $\phi_0$.  At every odd order in $\phi_0$, the metric backreacts on the scalar, and in general a new parameter from the potential appears.  For instance, at $\mathcal{O}(\phi_0^3)$, the coefficient of the $\phi^4$ term in a Taylor expansion of $V$ about $\phi=0$ becomes important.  At every even order in $\phi_0$, the scalar backreacts on the metric, and in general the black hole horizon shifts.  Call $r=R_h$ the location of the horizon in the full backreacted geometry.  $R_h$ can also be expanded in powers of $\phi_0$.

Since the scalar is now nonzero, we have added the term $\Lambda^{4-\Delta}\mathcal{O}_{\phi}$ to the field theory lagrangian.  The energy scale $\Lambda$ also appears in the $r^{\Delta-4}$ term of $\phi$ as $r\rightarrow\infty$:
\be
\label{phinearbound}
\phi(r) = (\Lambda L)^{4-\Delta}r^{\Delta-4} + \ldots .
\ee
Expanding Eq.~(\ref{firstscalar}) about $r=\infty$, we have
\be
\label{phinearbound2}
\phi(r) = \phi_0 \left(\frac{r}{r_h}\right)^{\Delta-4}\frac{\Gamma(\Delta/2-1)}{\Gamma(\Delta/4)^2} + \ldots .
\ee
We fix $\Lambda$ by setting $\Lambda L=1$.  Comparing Eqs.~(\ref{phinearbound}) and (\ref{phinearbound2}) then yields a relationship connecting $r_h$ and $\phi_0$:
\be
\label{rhandphi0}
r_h^{\Delta-4}=\phi_0\frac{\Gamma(\Delta/2-1)}{\Gamma(\Delta/4)^2}.
\ee
So $\phi_0$ is small when $r_h$ is large, and we are indeed performing an expansion valid at large $s$ and $T$.

The total differential order of the system Eqs.~(\ref{theeoms}) is four, so we must impose four boundary conditions at each order in $\phi_0$.  Three of them are to
\begin{itemize}
\item{maintain $\Lambda L=1$ and consequently Eq.~(\ref{rhandphi0})},
\item{preserve the boundary asymptotic $b\rightarrow r/L$ and consequently $h\rightarrow 1$, and}
\item{ensure that $\phi$ remains regular at $r=R_h$.}
\end{itemize}
The final boundary condition encodes the physical meaning of $\phi_0$.  To see this, imagine fixing a temperature-dependent observable $\Omega$ of $AdS_{5}$-Schwarzschild at some $\Omega_0$.  $\Omega$ could be a transport coefficient, the energy density, the entropy, or the temperature itself.  What form does the background take at $\Omega=\Omega_0$ when $\phi$ is turned on?  Our expansion should answer this question, so the fourth boundary condition enforces $\Omega=\Omega_0$ at every order in $\phi_0$.  

For example, if $\Omega=T$, we require that the temperature $T=|a(R_h)b(R_h)h'(R_h)|/4\pi$ remain at $r_h/\pi L$, its value in $AdS_{5}$-Schwarzschild.  Eliminating $r_h$ in favor of $T$ in Eq.~(\ref{rhandphi0}) gives 
\be
\phi_0=\frac{\Gamma(\Delta/4)^2}{\Gamma(\Delta/2-1)}\left(\pi LT\right)^{\Delta-4} , \nnb
\ee
so the expansion is essentially in the smallness of $1/LT$.

The expansion pursued here takes $\Omega=s$.  This is an especially convenient choice.  Our final boundary condition is then that the horizon of the black hole remains at $r=r_h$, so $h(r_h)=0$ at each order in $\phi_0$.  From Eqs.~(\ref{adsschwarzsandt}) and (\ref{rhandphi0}), we see that our expansion is in powers of
\be\label{phi0choice}
\phi_0=\frac{\Gamma(\Delta/4)^2}{\Gamma(\Delta/2-1)}\left(\frac{s\kappa_5^2}{2\pi}\right)^{(\Delta-4)/3}.
\ee
This is roughly an expansion in the smallness of $1/s\kappa_5^2$.

With an expansion in hand, it is possible to compute any desired transport coefficients to any desired order in $\phi_0$.  One must work to $\mathcal{O}(\phi_0^2)$ to determine the leading nontrivial high-temperature behavior of the transport coefficients.  At this order, the transport coefficients $\xi_i$ are only sensitive to the small $\phi$ part of the potential shown in Eq.~(\ref{asympV}).  Thus, they only depend on $\Delta$, and they take the form   
\be
\xi_i = \xi_i^{CFT} + C_{\xi_i}(\Delta) \phi_0^{2} +\mathcal{O}(\phi_0^4) \;. 
\ee
This is the reason that the normalized transport coefficients behave universally at high temperatures. Working to $\mathcal{O}(\phi_0^2)$, we have obtained closed-form solutions for $\phi(r)$, $b(r)$, and $h(r)$, but the expressions are cumbersome and we relegate them to an appendix.

\section{Transport coefficients}
\label{sec:TransportCoeffs}

\subsection{Speed of sound}

In this section, we briefly review the calculation of the speed of sound in single-scalar models~\cite{vs2CCN}.  The speed of sound $v_s^2$ can be computed from examining poles in Green's functions of the dual field theory's stress-energy tensor, but it can also be determined from the equation of state of a system~\cite{N2ThermoSound, D3D7sound, CascadingSound}.  We pursue this second way of calculating $v_s^2$.

The speed of sound can be obtained from $v_s^2 = d p /d \epsilon$, where $p$ is the pressure of a system and $\epsilon$ is its energy density.  For systems at zero chemical potential, this can be re-expressed as
\be
\label{vs2Formula}
v_s^2 = \frac{d \log T}{d \log s} \;.
\ee

Using the $\mathcal{O}(\phi_0^2)$ high-temperature expansion described in the previous section, we can easily determine the first nontrivial correction to $v_s^2$.  Evaluating Eq.~(\ref{SandT}) and applying Eq.~(\ref{vs2Formula}), we arrive at
\bea
\label{vs2_result}
 v_{s}^{2} (\phi_0) &=& 1/3 - C_{v_s^2}(\Delta) \phi_{0}^{2}+\mathcal{O}(\phi_{0}^{3}),\; \mathrm{where}\nnb \\
 C_{v_s^2}(\Delta) &=& \frac{1}{576} (\Delta-4)^2 \Delta \Big[16+(\Delta-4)\Delta  \nnb\\ 
 &\times& \, \int_{1}^{\infty} {ds\,s\, {}_{2}F_{1}(2-\Delta/4, 1+\Delta/4; 2;1 - s)^2} \,\Big] \nnb \\
&=& \frac{1}{9 \pi} (4-\Delta)(2-\Delta) \tan \left(\pi \Delta/4\right)\;,
\eea
where the simplified form in the last line can be found using some standard properties of Meijer G-functions.  

We note two important implications of the above result.  First, since $C_{v_s^2}(\Delta)$ is positive for $2 < \Delta < 4$, the speed of sound is always bounded from above by $1/3$ at high temperatures in single-scalar models \cite{vs2CCN,HohlerStephanov}.  Second, to the order to which we are working,
\be
\label{TfromPhi0}
\phi_0 = (\pi L T)^{\Delta-4} \frac{\Gamma(\Delta/4)^2}{\Gamma(\Delta/2-1)} \;,
\ee
so the temperature dependence of $1/3 - v_s^2$ takes the simple universal form $(1/3 - v_s^2) \sim (\pi L T)^{2(\Delta-4)}$ as $T\rightarrow \infty$.  Below, we will see that this feature also applies to the other transport coefficients.

\subsection{Conductivity}  

Suppose the dual field theory contains a conserved $U(1)$ charge and an associated conserved current $J_{\mu}$, where $\mu = t, x, y, z$.  Then there are transport coefficients associated with these charges: the DC conductivity $\sigma$, the charge susceptibility $\Xi$, and the diffusion coefficient $D$.  In this section, we calculate $\sigma$~\cite{SigmaCalculations} at high temperatures.  

In the gravity dual, there is a $U(1)$ gauge field $A_M$ ($M = \mu, r$) that is dual to the conserved current in the field theory, and we must add the term 
\be
- \int{d^{5}x\,\sqrt{-g} \frac{1}{4 g_5^2} F_{MN}F^{MN}}
\ee
to the action Eq.~(\ref{S5D}), where $g_5^2$ is the $5D$ gauge coupling and $F_{MN}$ is the field strength.  In terms of the parameters of the dual field theory, $g_5^2/L \sim \Nc^{-2}$.   We work in the probe limit, where the bulk $U(1)$ gauge field does not backreact on the metric.  The DC conductivity $\sigma$ is
\be
\sigma \delta^{ij} = \lim_{\omega \rightarrow 0} \frac{G_R^{ij} (\omega, \mathbf{0})}{i \omega} ,
\ee
where $G_R^{ij}(\omega,\mathbf{k})$ is the retarded Green's function of $J_{i}(\omega, \mathbf{k})$ ($i = x,y,z$), and $\omega$ is the frequency.  This Green's function can be calculated via the gauge/gravity duality.  Ref.~\cite{IqbalLiu} showed that in the low-frequency limit, $\sigma$ can be extracted from the geometry of the gravity dual at the horizon of the black hole via the membrane paradigm \cite{MembraneStuff}:
\be
\label{sigmaFormula}
\sigma =\left. \frac{1}{g_5^2} \frac{b(r)}{a(r)^{3}} \sqrt{-g(r)} \; \right|_{r = r_h} \;.
\ee

Using the metric coefficients given in the appendix together with Eqs.~(\ref{TfromPhi0}) and~(\ref{sigmaFormula}), we find that at high temperatures, the conductivity $\sigma$ in gauge theories with single-scalar gravity duals is
\bea
\label{sigmaResult}
\tilde{\sigma}(T) &\equiv& \frac{\sigma g_5^2}{\pi L T} = 1 - C_{\tilde{\sigma}}(\Delta) \phi_0^2 \,,\; \mathrm{where} \\
C_{\tilde{\sigma}}(\Delta) &=&  \frac{1}{6\pi} (2-\Delta) \tan\left(\frac{\pi \Delta}{4}\right). \nnb
\eea

\begin{figure}[t]
\centering
\includegraphics[scale=1]{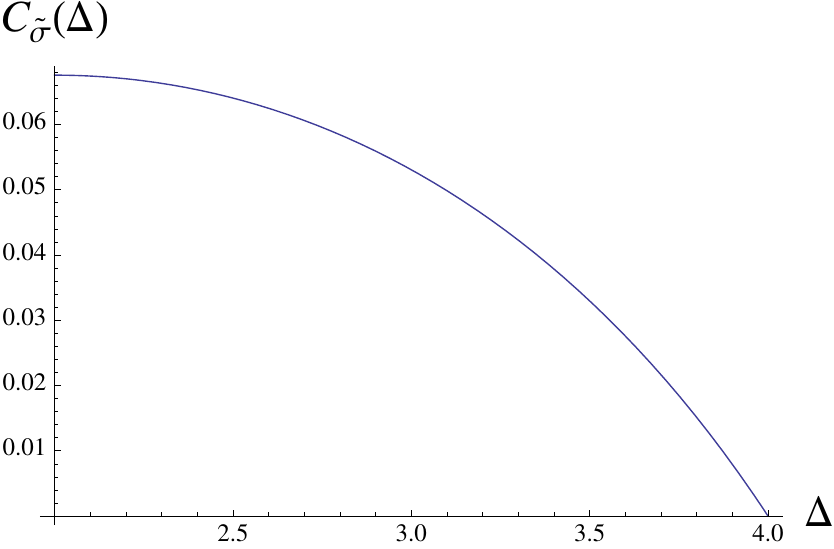}
\caption{Plot of $C_{\tilde{\sigma}}$ versus $\Delta$.}
\label{SigmaPlot}
\end{figure}

A plot of $C_{\tilde{\sigma}}$ is given in Fig.~\ref{SigmaPlot}.  Note that like $v_s^2$, $\tilde{\sigma}$ is bounded from above at high temperatures.  This time, the limiting value is $1$.  The temperature dependence of $\tilde{\sigma}$ also takes the same form as that of $v_s^2$, since  $(1 - \tilde{\sigma}) \sim T^{2(\Delta-4)}$ in the high temperature limit.  From the expressions above, we see that the transport coefficient ratio Eq.~(\ref{universal_relation}) is satisfied, and in this class of models we have
\be
R_{\tilde{\sigma},v_s^2} (\Delta) = \lim_{T\rightarrow \infty} \frac{\partial \xi_i /\partial \log T}{\partial \xi_j /\partial \log T}  = \frac{3}{8-2\Delta}\;.
\ee
As advertised, $R_{\tilde{\sigma},v_s^2}(\Delta)$ is a temperature-independent constant.

\subsection{Diffusion coefficient and charge susceptibility}

We can also calculate the diffusion coefficient associated with the $U(1)$ charges.  The diffusion coefficient $D$ describes the relaxation to equilibrium of a small charge density perturbation in (for definiteness) the $z$ direction in the field theory.  $D$ appears as a pole in the two-point retarded Green's function for $J_z$ at $\omega(k) =  i D k^2$.   This correlation function can again be calculated using the gauge/gravity duality.  For the class of theories we are considering, it was shown in Ref.~\cite{IqbalLiu} that $D$ can be extracted directly from the geometry through the simple formula
\be
D = \sigma \int_{r_h}^{\infty} {dr \frac{a(r)^2}{b(r)^2 \sqrt{-g}} g_5^2}   \;.
\ee

Using the high temperature expansion discussed in Sec.~\ref{sec:HighT}, we can simply read off $D$ from the metric components.  To $\mathcal{O}(\phi_0^2)$,
\begin{widetext}
\bea
\label{Dresult}
\tilde{D}(T) &\equiv& 2 \pi T D = 1+ C_{\tilde{D}}\phi_0^2 \,,\; \mathrm{where} \\
C_{\tilde{D}} &=& \frac{1}{96\pi} \Bigg(4\pi \Delta(\Delta-4) - 32(\Delta-2)\tan(\pi \Delta/4) 
+\pi \Delta(\Delta-4) \int_{1}^{\infty} {du\, u^5 {}_{2}F_{1}}(2-\frac{\Delta}{4}, 1+\frac{\Delta}{4}; 2; 1-u^4)^2  \Bigg). \nnb
\eea
\end{widetext}

\begin{figure}[t]
\centering
\includegraphics[scale=1]{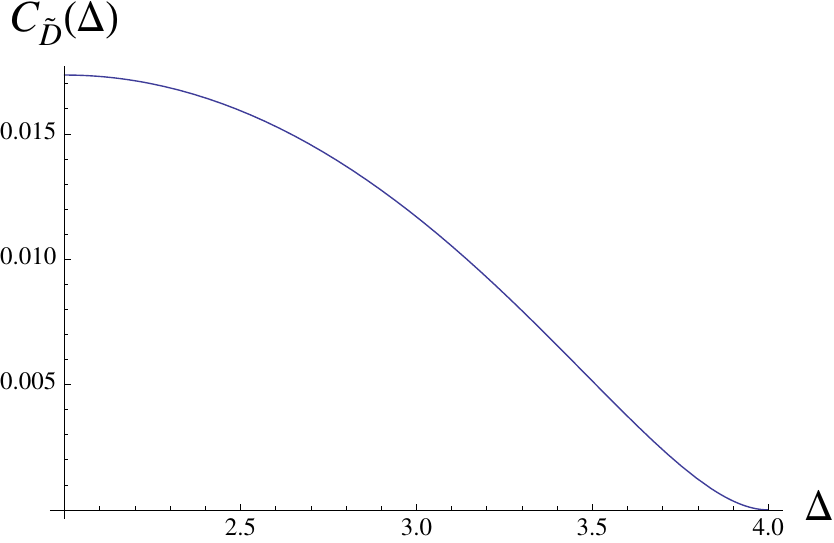}
\caption{Plot of $C_{\tilde{D}}$ versus $\Delta$.}
\label{DPlot}
\end{figure}

A plot of $C_{\tilde{D}}$ is given in Fig.~\ref{DPlot}.  At high temperatures, $2\pi T D \geq 1$.  Defining $\xi_1 = \tilde{DT}$ and $\xi_2 = v_s^2$, we see that as with the conductivity, the transport coefficient ratio Eq.~(\ref{universal_relation}) is satisfied.

In theories with gravity duals of the sort that we are considering here, it is known that $D$ and $\sigma$ satisfy the Einstein relation $D \Xi = \sigma$, where $\Xi$ is the charge susceptibility~\cite{IqbalLiu}.  Using Eqs.~(\ref{sigmaResult}) and (\ref{Dresult}), we have
\be
\label{XiResult}
\tilde{\Xi} \equiv \frac{g_5^2 \Xi}{2L(\pi T)^2} = 1- (C_{\tilde{\sigma}}+C_{\tilde{D}}) \phi_{0}^{2} \;.
\ee
Since $C_{\tilde{\sigma}}$ and $C_{\tilde{D}}$ are positive, $\tilde{\Xi}\le 1$ at high temperatures.

Kovtun and Ritz have proposed a bound on $D = \sigma/\Xi$ in systems with gravity duals \cite{KovtunRitz}: 
\be
D \geq \frac{1}{2\pi T} .
\ee
Our result is consistent with this proposal.


\subsection{Bulk viscosity} 

The bulk viscosity $\zeta$ characterizes the resistance of a fluid to volume change under external stress.  It can be extracted from the low-frequency behavior of an $SO(3)$-symmetric correlation function of the stress-energy tensor.  In a holographic dual, this correlation function is determined from $SO(3)$-invariant time-dependent perturbations of the metric~\cite{bulklitany,BenincasaEtAZeta,BuchelPagnuttiZeta}.  Calculations of $\zeta$ are usually challenging because the relevant metric perturbations mix with perturbations of the other bulk fields.  However, an elegant formalism that applies to single-scalar duals was developed in Ref.~\cite{stevebulk}.  A key idea is to use a gauge where the scalar $\phi$ is itself the holographic coordinate: $r=\phi$ in Eq.~(\ref{metricansatz}).  In this gauge, the perturbations decouple, making calculations of $\zeta$ much easier than in the gauge we use in the rest of this paper.  Thus, for much of this section, we employ the $r=\phi$ gauge.

As shown in Ref.~\cite{stevebulk}, to extract $\zeta$ in the $r=\phi$ gauge it is sufficient to consider the diagonal metric ansatz
\bea
\label{zetametric}
ds^2 &=&g_{00}dt^2+g_{11}d\vec{x}^2+g_{55}d\phi^2 ,\;  \mathrm{where} \\
g_{00}&=&-e^{2A}h(1+\frac{\lambda}{2}H_{00})^2 \;,\; g_{11}=e^{2A}(1+\frac{\lambda}{2}H_{11})^2\nnb\\
g_{55}&=&\frac{e^{2B}}{h}(1+\frac{\lambda}{2}H_{55})^2\nnb \;.
\eea
Above, $A$, $h$, and $B$ are functions of only $\phi$, $H_{00}$, $H_{11}$, and $H_{55}$ are functions of both $t$ and $\phi$, and $\lambda$ is a formal expansion parameter.  $H_{00}$, $H_{11}$ and $H_{55}$ parameterize the $SO(3)$-invariant metric perturbations.  At zeroth order in $\lambda$, the Einstein equations determine the geometry around which we are perturbing.  In our case, this is a geometry obtained from the high-temperature expansion described in a previous section.  At first order in $\lambda$, the Einstein equations yield a system that determines $H_{00}$, $H_{11}$, and $H_{55}$.  Assuming $H_{11}(t,\phi)=e^{-i\omega t}h_{11}(\phi)$, the $x_{1}x_{1}$ component of the Einstein equations gives
\bea
\label{bulkeqphi}
h_{11}''= \left(-\frac{1}{3A'}-4A'+3B'-\frac{h'}{h} \right)h_{11}'\nnb\\
+\left(-\frac{e^{-2A+2B}}{h^2}\omega^2+\frac{h'}{6hA'}-\frac{h'B'}{h}\right)h_{11} \;,
\eea
where the primes denote derivatives with respect to $\phi$.  As shown in Ref.~\cite{stevebulk}, for $\omega=0$ in Eq.~(\ref{bulkeqphi}), $h_{11}$ can be related to the bulk viscosity.  With the boundary conditions $h_{11}\rightarrow 1$ as $\phi\rightarrow 0$ and $h_{11}$ is regular at $\phi=\phi_{H}$, one finds that
\be
\label{bulkformula}
\frac{\zeta}{s}=\frac{1}{4\pi}h_{11}(\phi_H)^2\frac{V'(\phi_H)^2}{V(\phi_H)^2} \;,
\ee
where $s$ is the entropy density and $\phi_H$ is the value of the scalar at the horizon.

Our high-temperature expansion for the background perturbs around $AdS$-Schwarzschild, for which $\phi=0$ everywhere.  Unfortunately, this makes $\phi$ an inappropriate radial variable, but we can still use the equation for $h_{11}$.  First we change coordinates in Eq.~(\ref{bulkeqphi}) (with $\omega=0$) from the $r=\phi$ gauge back to the $a=r$ gauge.  Then we plug in expressions for $b$, $h$, and $\phi$ up to second order in $\phi_0$.  Defining $u=r/r_h$ and retaining only the leading term in powers of $\phi_0$ in the coefficients of each of $h_{11}$, $dh_{11}/du$, and $d^{2}h_{11}/du^{2}$, we arrive at
\be
\label{h11fineqn}
h_{11}''=\alpha(u, \Delta)h_{11}'+\beta(u, \Delta)h_{11} \;.
\ee
Here, primes denote derivatives with respect to $u$, and the coefficients $\alpha(u,\Delta)$ and $\beta(u, \Delta)$ are given in the appendix.  $\alpha(u, \Delta)$ and $\beta(u, \Delta)$ do not depend on $r_h$, $\phi_0$, or $\phi_H$.  Therefore, Eq.~(\ref{h11fineqn}) does not involve $r_h$, $\phi_0$, or $\phi_H$ --- the only parameters that depend on temperature.  So at high temperatures, $h_{11}(\phi=\phi_H)=h_{11}(u=1)$ depends only on $\Delta$; all of the temperature dependence of $\zeta/s$ is in the $V'(\phi_H)^2/V(\phi_H)^2$ factor of Eq.~(\ref{bulkformula}).  Plugging $r=r_h$ into Eq.~(\ref{firstscalar}), we see that $\phi_H = \phi_0$ up to $\mathcal{O}(\phi_0^2)$.  It follows that
\be
\label{ZetaResult}
\frac{\zeta}{s} = C_{\zeta/s}(\Delta) \phi_0^2 \;,
\ee
where $C_{\zeta/s}(\Delta)$ can be determined from the solution to Eq.~(\ref{h11fineqn}).  The form of Eq.~(\ref{ZetaResult}) implies that at high temperatures, the temperature dependence of $\zeta/s$ takes the form $\zeta/s \sim T^{-2(4 - \Delta)}$, as it does for the other transport coefficients we consider.  The universal relation Eq.~(\ref{universal_relation}) is thus satisfied for $\zeta/s$.

\begin{figure}[t]
\centering
\includegraphics[scale=1]{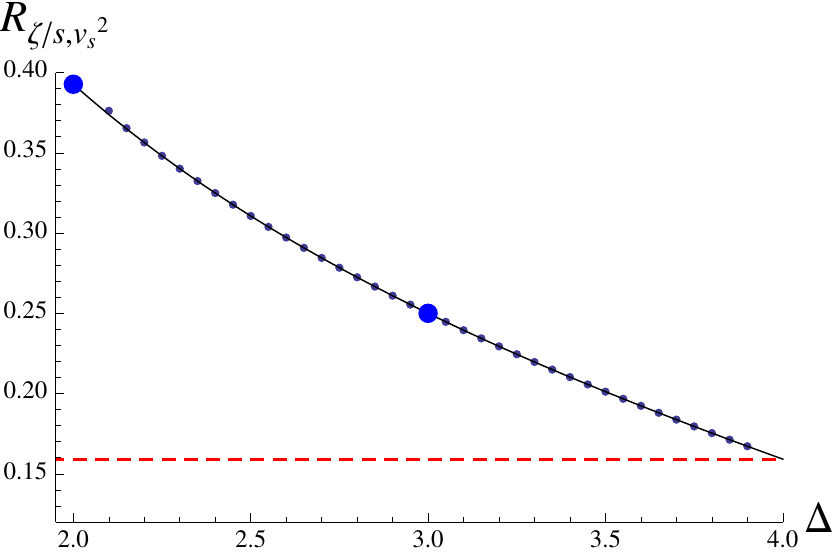}
\caption{Plot of $R_{\zeta/s, v_s^2} (\Delta)=\frac{\partial (\zeta/s) /\partial \log T}{\partial (v_s^2) /\partial \log T}$ vs. $\Delta$.  The small dots were obtained from a numerical solution of Eq.~(\ref{h11fineqn}), which becomes unreliable near $\Delta=2$.  The solid curve represents a guess at an analytical form described in the text,  and the two large dots are taken from the $\zeta/\eta$ results of Refs.~\cite{BenincasaEtAZeta,BuchelPagnuttiZeta}.   The dashed line is at $R_{\zeta/s, v_s^2}=1/2\pi$, the bound suggested in Ref.~\cite{BuchelZetaBound}.
}
\label{ZetaPlot}
\end{figure}

There is no obvious analytical solution to Eq.~(\ref{h11fineqn}), so we solve it numerically and subsequently extract $\zeta/s$.  In Fig.~\ref{ZetaPlot}, we plot the relevant quantity from  Eq.~(\ref{universal_relation}): 
\be
\label{forZetaPlot}
R_{\zeta/s, v_s^2} (\Delta) \equiv\frac{\partial (\zeta/s) /\partial \log T}{\partial (v_s^2) /\partial \log T}.
\ee
The dots denote numerical results.  We note that a simple analytic formula
\be
\label{ZetaAnalytic}
C_{\zeta/s}(\Delta)=\frac{1}{9}(\Delta-4)^2
\ee  
fits our numerical results to a remarkable degree of precision except near $\Delta=2$, where our numerics should not be trusted.   For instance, at $\Delta=3$, the analytical formula predicts $C_{\zeta/s}(\Delta=3)=1/9$, and the numerical result is $0.1111111$.   

Furthermore, we can compare this conjectured analytic formula for $\zeta/s$ against the high-temperature calculations done for the $\mathcal{N}=2^*$ theory in Refs.~\cite{BenincasaEtAZeta,BuchelPagnuttiZeta}.  $\mathcal{N}=2^*$ theory is a deformation of $\mathcal{N}=4$ SYM that is obtained by turning on masses for the bosons and/or fermions in two of the $\mathcal{N}=1$ chiral multiplets that are part of the $\mathcal{N}=4$ gauge theory.  In the notation used in this paper, this corresponds to looking at bulk scalars with $\Delta=2$ or $\Delta=3$.  For $\Delta=3$, corresponding to turning on masses for fermions, we find that $R_{\zeta/s, v_s^2} (\Delta = 3) = \pi$, while for $\Delta \rightarrow 2$, corresponding to turning on masses for the bosons, we get $R_{\zeta/s, v_s^2} (\Delta \rightarrow 2) = \pi^{2}/2$.  These results agree with the results of Ref.~\cite{BuchelPagnuttiZeta}.

Given the remarkable agreement of Eq.~(\ref{ZetaAnalytic}) with our numerical solutions and with the results of Refs.~\cite{BenincasaEtAZeta,BuchelPagnuttiZeta}, it seems likely that an analytical high-temperature expression for $\zeta/s$ exists and is given by Eqs.~(\ref{ZetaResult}),~(\ref{ZetaAnalytic}).  Unfortunately, we have not been able to show this directly from Eq.~(\ref{h11fineqn}), and an analytical demonstration that Eq.~(\ref{ZetaAnalytic}) gives the correct expression for $\zeta/s$ at high temperatures is an important subject for future work. 

It was proposed in Ref.~\cite{BuchelZetaBound} that
\be
\frac{\zeta}{\eta} \ge 2 \left(\frac{1}{3} - v_s^2\right) 
\ee
in all theories with gravity duals.  This translates to the bound $R_{\zeta/s, v_s^2} \geq 1/2\pi$.  As can be seen from Fig.~\ref{ZetaPlot}, the bound is always satisfied at high temperatures in the class of theories we consider for $2 < \Delta < 4$ and becomes saturated as $\Delta \rightarrow 4$.

\section{Systems with multiple scalars}  
\label{sec:MultiScalar}

Let us now treat the generalization of our results to theories with gravity duals sourced by more than one scalar.  The discussion below applies to all of the normalized transport coefficients we have computed except for the bulk viscosity $\zeta/s$.  Our treatment of $\zeta/s$ above used the $r=\phi$ gauge to decouple the metric and scalar perturbations, but it is not clear this approach generalizes straightforwardly to systems with multiple scalars.  We leave the investigation of $\zeta/s$ in gravity duals with multiple scalars for future work.

To be specific, we examine the two-scalar case, since the generalization to an arbitrary number of scalars is then easy to see.  The action of the gravity dual with two scalars is
\be
\label{S5Dmulti}
S = \frac{1}{2 \kappa_{5}^2} \int{d^{5}x \sqrt{-g} \left[ R - \frac{1}{2}(\partial \phi)^2 - \frac{1}{2}(\partial \chi)^2 - V(\phi, \chi) \right]} \;,
\ee
where $\phi$ and $\xi$ are the two scalar fields, and $V(\phi, \chi)$ is an analytic potential that is even in both $\phi$ and $\chi$ and satisfies $V(\phi=0, \chi=0) < 0$.  We focus on field theories that can be viewed as CFTs deformed by the addition of two relevant operators $\mathcal{O}_{\phi}$, $\mathcal{O}_{\chi}$.  In the gravity theory, this corresponds to the restriction that
\bea
\label{asympVmulti}
\lim_{r\rightarrow \infty} V(\phi) &=& -\frac{12}{L^2} + \frac{1}{2L^2} \Delta_{\phi}(\Delta_{\phi}-4) \phi^2 + \mathcal{O}(\phi^{4}) \nnb \\
&+& \frac{1}{2L^2} \Delta_{\chi}(\Delta_{\chi}-4) \chi^2  +  \mathcal{O}(\chi^{4}) \;,
\eea
where $\Delta_{\phi}$ and $\Delta_{\chi}$ are the UV scaling dimensions of $\mathcal{O}_{\phi}$ and $\mathcal{O}_{\chi}$, respectively.  Just as in the single-scalar case, we restrict our attention to $2 < \Delta_{\phi},\Delta_{\chi} <4$.

The high-temperature expansion can be developed in the same way as for a single scalar.  The difference is that there are now two expansion parameters, $\phi_0$ and $\chi_0$, and two energy scales associated with the CFT deformations, $\Lambda_{\phi}$ and $\Lambda_{\chi}$~\footnote{The boundary asymptotics of the second scalar $\chi$ could be arranged so the CFT lagrangian is not deformed.  This setup corresponds to a spontaneously broken $\mathbb{Z}_2 \times \mathbb{Z}_2$ symmetry at some finite transition temperature $T_c$~\cite{PTnearBHH}.  At arbitrarily high temperatures, however, $\chi$ is not active and the system behaves just like a single-scalar system.}.   Holding $\Lambda_{\chi}$ and $\Lambda_{\phi}$ fixed gives relationships that connect $r_h$ and $\chi_0$ as well as $r_h$ and $\phi_0$:
\bea
\label{rhandphi0chi0}
r_h^{\Delta_{\phi}-4}&=&\phi_0\frac{\Gamma(\Delta_{\phi}/2-1)}{\Gamma(\Delta_{\phi}/4)^2} \\
r_h^{\Delta_{\chi}-4}&=& \chi_0 \left(\frac{\Lambda_{\chi}}{\Lambda_{\phi}}\right)^{\Delta_{\chi}-4} \frac{\Gamma(\Delta_{\chi}/2-1)}{\Gamma(\Delta_{\chi}/4)^2} . \nnb
\eea
This is the only substantive change from the single-scalar case.

To calculate the leading correction to the high-temperature behavior of transport coefficients, we need to work to order $\mathcal{O}(\phi_0^2, \chi_0^2)$.  For a given normalized transport coefficient $\xi_i$,
\be
\xi_i = \xi_i^{CFT} + C_{\xi_i,\phi}(\Delta_{\phi}) \phi_0^{2} + C_{\xi_i,\chi}(\Delta_{\chi}) \chi_0^{2} +\mathcal{O}(\phi_0^4,\chi_0^4,\phi_{0}^{2}\chi_0^2) \;. \nnb
\ee
Note that terms proportional to $\phi_0\chi_0$ cannot arise because there are no terms proportional to $\phi\chi$ in a Taylor expansion of $V$ about $(\phi,\chi)=(0,0)$.  

Consider the case $\Delta_{\phi} \neq \Delta_{\chi}$, and suppose without loss of generality that $\Delta \equiv \Delta_{\phi} > \Delta_{\chi}$.  From Eqs.~(\ref{adsschwarzsandt}),~(\ref{rhandphi0chi0}), we have that $\phi_0^2 \gg \chi_0^2$ as $T\rightarrow \infty$.  So the leading deviation from the $T\rightarrow\infty$ behavior is driven by the least relevant operator, and
\be
\xi = \xi_i^{CFT} + C_{\xi_i,\phi}(\Delta) \phi_0^{2} \nnb
\ee
plus terms that vanish as $T\rightarrow \infty$.  Therefore, Eq.~(\ref{universal_relation}) holds when $\Delta_{\phi} > \Delta_{\chi}$. 

Now consider the case $\Delta = \Delta_{\phi} = \Delta_{\chi}$.  Using Eq.~(\ref{TfromPhi0}), we can write
\be
\label{EqualDelta}
\xi = \xi_i^{CFT} + C_{\xi_i} (\Delta) \frac{\Gamma(\Delta/4)^4}{\Gamma(\Delta/2-1)^2} \gamma_{\phi, \chi} T^{-2 (4-\Delta)}\;,
\ee
where
\be
\gamma_{\phi,\chi} = \Lambda_{\phi}^{2(4-\Delta)}+\Lambda_{\chi}^{2(4-\Delta)} \;.
\ee
From Eq.~(\ref{EqualDelta}), it is clear that $\gamma_{\phi,\chi}$ will cancel in the ratio in Eq.~(\ref{universal_relation}).  Thus, Eq.~(\ref{universal_relation}) continues to hold for any $\Delta_{\phi}$ and $\Delta_{\chi}$.  

The analysis above is trivially extended to systems with $n$ scalars.  The essential idea is that high temperatures pick out only the least relevant deformations of the CFT, so only the largest $\Delta_i$ matters.  Therefore, the results in this paper (except the ones involving the bulk viscosity, which needs a separate treatment) apply to theories with $n$ scalar fields dual to $n$ relevant deformations.  For instance, the sound bound $v_s^2 \le 1/3$ holds at high temperatures in theories with gravity duals containing multiple scalars.


\section{Conclusions}
\label{sec:Conclusions}

In this paper, we have used the high-temperature expansion developed in Ref.~\cite{vs2CCN} to show that there are universal relations between dimensionless combinations of transport coefficients $\xi_i \in \{\eta/s,  v_s^2, 2 \pi T D, \sigma/T, \Xi \}$ in theories with gravity duals sourced by one or more scalar fields.   We have also shown that $\zeta/s$ has the same universal behavior as the other transport coefficients in theories with single-scalar gravity duals.

We have shown that the ratio of the derivatives of any two dimensionless transport coefficients $\xi_1, \xi_2$ becomes a temperature-independent nonzero constant that depends only on $\Delta = \max(\Delta_i)$ as $T\rightarrow \infty$:
\be
R_{\xi_{i}, \xi_j}(\Delta) =  \lim_{T\rightarrow \infty} \frac{d \xi_i}{d \xi_j } =  \lim_{T\rightarrow \infty} \frac{\partial \xi_i /\partial \log T}{\partial \xi_j /\partial \log T}  .
\ee
To understand the utility of this result, suppose that one has a system described by the class of gravity duals considered here, and suppose one could measure a transport coefficient at high temperatures.  Then one could determine $\Delta$.  (In a multiscalar case, using the theoretical expression for $C_{\xi_i}$, one could also infer $\gamma_{\phi,\chi,\ldots}$ from this measurement.)  The high-temperature behavior of \emph{all} of the other transport coefficients would then be a prediction of the theory.

We have shown that the sound bound proposed in Refs.~\cite{vs2CCN,HohlerStephanov} for single-scalar systems also holds in systems with multiple scalars.   A calculation of the bulk viscosity for single-scalar systems showed that $\eta/s$ has the same temperature scaling as the other observables.  The bound on $\zeta/\eta$ of Ref.~\cite{BuchelZetaBound} has been shown to hold in single-scalar systems at high temperatures for $2< \Delta < 4$, with saturation as $\Delta \rightarrow 4$.  

We have also explored the transport coefficients of $U(1)$ charges in the strongly coupled gauge theory plasmas described by gravity duals with multiple scalars, and have seen that there are high-temperature bounds on $\sigma$, $D$ and $\Xi$.  The bound we observed on $\sigma/\Xi = D$ is consistent with the one proposed in Ref.~\cite{KovtunRitz}.   It is not clear how generic the bound on $D$ is, since it has been observed that $D$ can fall below the conjectured bound in the D3/D7 system at high temperatures~\cite{D3D7SmallD}.

A natural question for future work is whether the existence of relations like Eq.~(\ref{universal_relation}) is a generic feature of systems with gravity duals.  If these relations do indeed hold generically in a broader class of systems with gravity duals, it would be interesting to investigate the finite $\lambda$ and $N_c$ corrections to the high-temperature behavior of transport coefficients in field theories with string duals to see whether Eq.~(\ref{universal_relation}) continues to hold.

{\it Acknowledgements.}  We thank Tom Cohen for encouraging us to write up these observations, very helpful conversations, and collaboration on a related project.  We also thank Steve Gubser, Paul Hohler, Silviu Pufu, F\'abio Rocha, Misha Stephanov, and Amos Yarom for helpful discussions.  A.~C. thanks the US DOE for support under grant DE-FG02-93ER-40762, and A.~N. thanks the US NSF for support under grant PHY-065278.

\section{Appendix}  
We work at a fixed entropy, corresponding to fixing the location of the black hole horizon at its $AdS_5$-Schwarzschild value, $r= r_h$. A calculation of the back-reaction of one scalar field on the geometry of the gravity dual gives the following results for $h(r)$ and $b(r)$ to $\mathcal{O}(\phi_0^2)$:
\begin{widetext}
\bea
b(r) &=& \frac{r}{L} - \frac{\Delta^2(\Delta-4)^2 \phi_0^2 r }{96 L r_h^8} \int_{\infty}^{r} {dx \, x^7 \left({}_{2} F_{1}(2-\frac{\Delta}{4}, 1+\frac{\Delta}{4};2; 1- \frac{x^4}{r_h^4}) \right)^2} \\
h(r) &=& 1 - \frac{r_h^4}{r^4} -\frac{r_h \phi_0^2}{6 r^4} \int_{r_h}^{r}{dx\, x^2 f(x)}, \; \mathrm{where} \\
\frac{16 r_h^9}{\Delta(\Delta-4) x} f(x) &=&  16 r_h^8\, {}_{2} F_{1}(1-\frac{\Delta}{4}, \frac{\Delta}{4}; 1; 1- \frac{x^4}{r_h^4})+x^4(r_h^2-x^4) \Delta(\Delta-4)\,{}_{2} F_{1}(2-\frac{\Delta}{4}, 1+\frac{\Delta}{4};2; 1- \frac{x^4}{r_h^4}) \nnb\\
&-& 8\Delta(\Delta-4) \int_{\infty}^{x} {dy \, y^7 \left({}_{2} F_{1}(2-\frac{\Delta}{4}, 1+\frac{\Delta}{4};2; 1- \frac{y^4}{r_h^4}) \right)^2}.  \nnb
\eea

The coefficient functions appearing in Eq.~(\ref{h11fineqn}), which are used in the calculation of $\zeta/s$, are given by:  
\bea
\alpha(u,\Delta) &=& \frac{14 u^4 - 9}{u-u^5} - \frac{u^3 \, (\Delta-8)(4+\Delta) \, {}_{2} F_{1}(3-\frac{\Delta}{4}, 2+\frac{\Delta}{4}; 3; 1- u^4)}{4 \, {}_{2} F_{1}(2-\frac{\Delta}{4}, 1+\frac{\Delta}{4};2; 1- u^4)} \\
\beta(u,\Delta) &=& \frac{32 \, {}_{2} F_{1}(1-\frac{\Delta}{4}, \frac{\Delta}{4};2; 1- u^4)+(\Delta-8)(4+\Delta)\,  {}_{2} F_{1}(1-\frac{\Delta}{4}, \frac{\Delta}{4};3; 1- u^4)}{2 u^{6}(u^4-1)\,  {}_{2} F_{1}(2-\frac{\Delta}{4}, 1+\frac{\Delta}{4};2; 1- u^4)}.
\eea
\end{widetext}

\end{document}